\documentclass[]{fairmeta}

\usepackage{subcaption}

\usepackage{amsmath}
\usepackage{amsfonts}
\usepackage{amssymb}
\usepackage{soul}
\usepackage{xcolor}

\usepackage{natbib}
\usepackage{hyperref}

\usepackage{tcolorbox}
\tcbuselibrary{skins, breakable, theorems, listings}

\newtcblisting{promptbox}[1]{
  enhanced,
  colback=black!3!white,    
  colframe=black!30!white,  
  boxrule=1.5pt,
  arc=3pt,
  left=5pt,
  right=5pt,
  top=3pt,
  bottom=1pt,
  title={#1},
  fonttitle=\sffamily\bfseries\small,
  coltitle=white,
  attach boxed title to top left={xshift=10pt, yshift=-\tcboxedtitleheight/2},
  boxed title style={
    colback=black!70,       
    colframe=black!70,
    arc=0pt,
    outer arc=0pt,
  },
  listing only,
  float,
  listing options={
    basicstyle=\ttfamily\small, 
    breaklines=true,            
    columns=fullflexible
  }
}

\newtcblisting{promptboxhl}[1]{
  enhanced,
  colback=black!3!white,    
  colframe=black!30!white,  
  boxrule=1.5pt,
  arc=3pt,
  left=5pt,
  right=5pt,
  top=3pt,
  bottom=1pt,
  title={#1},
  fonttitle=\sffamily\bfseries\small,
  coltitle=white,
  attach boxed title to top left={xshift=10pt, yshift=-\tcboxedtitleheight/2},
  boxed title style={
    colback=black!70,       
    colframe=black!70,
    arc=0pt,
    outer arc=0pt,
  },
  listing only,
  float,
  listing options={
    basicstyle=\ttfamily\small, 
    breaklines=true,            
    columns=fullflexible,
    escapeinside={(*}{*)}
  }
}

\title{An LLM-powered Agentic Recommendation System for Connected TV Content Discovery}

\author[1]{Lei Shi}
\author[1]{Di Wang}
\author[1]{Harry Tran}
\author[1]{Helsing Xu}
\author[1]{Yuchen Lu}
\author[1]{Dhara Ghodasara}
\author[1]{Wilson Chaney}
\author[1]{Xueting Liao}
\author[1]{Guohao Xiao}
\author[1]{Jerry Yu}
\author[1]{Huayu Ding}
\author[1]{Reza Mirghaderi}
\author[1]{David Fan}
\author[1]{Qi Guo}
\author[1]{Chongguang He}
\author[1]{Warren Wang}
\author[1]{Warren Deng}
\author[1]{Mingze Gao}
\author[1]{Shike Mei}
\author[1]{Shuo Tang}
\author[1]{Zhe Zhang}
\author[1]{Jianming He}
\author[1]{Abhishek Kumar}
\author[1]{Haotian Wu}
\author[1]{Hamed Firooz}
\author[1]{Li Li}

\affiliation[1]{Meta}

\abstract{Recommendation systems, from traditional multi-stage to recent unified generative architectures, face challenges in incorporating diverse contextual signals, such as trending topics, breaking news, cultural events, and cross-surface user activities, into their ranking pipelines. These systems are designed to consume structured behavioral signals with consistent schemas, and lack the reasoning capability to naturally process unstructured or heterogeneously formatted contextual information. Incorporating such signals typically requires feature engineering, bespoke data pipelines, and carefully tuned heuristics. In this paper, we present an LLM-powered agentic recommendation system designed for Connected TV (CTV) content discovery that addresses these limitations. Our system leverages the reasoning capabilities of large language models to naturally process and synthesize diverse signals across varying schemas and structures, eliminating much of the manual integration inherent in traditional ranking and retrieval systems. Recognizing that current LLM-based solutions still fall short of traditional machine learning models in several recommendation tasks, including retrieval efficiency, personalization precision, and scalability, we adopt an agentic architecture that orchestrates specialized components, allowing each sub-task to be handled by the most suitable method, whether LLM-based or traditional ML. The main contribution of this work is our engineering approach to successfully overcoming the practical limitations of enabling LLM for recommendation, particularly inference latency. We share insights from our work and discuss the trade-offs and lessons learned in building a hybrid system that combines the flexibility of LLMs with the performance of established recommendation techniques.}

\date{\today}
\correspondence{Lei Shi at \email{leis@meta.com}, Li Li at \email{leeley@meta.com}}

\begin{document}

\maketitle

\section{Introduction}
Connected TV (CTV) has emerged as a growing surface for content consumption, offering a lean-back, large-screen experience that is distinct from mobile-centric environments. While conventional recommendation interfaces present users with a single ranked list of content, such as infinite-scroll news feeds or short-form video queues, the CTV home screen shows multiple horizontal carousels organized around coherent topics or themes. This architectural shift encourages users to explore content across diverse categories. Consequently, this paradigm transitions the recommendation objective from producing a single total ordering of items to curating multiple, semantically distinct content groupings. These groupings must maintain strict intra-group coherence while collectively maximizing coverage across both evergreen user interests and real-time trends.

Furthermore, the lean-back nature of television viewing shapes user expectations distinctly from mobile consumption. On CTV surfaces, users expect curated, timely, and contextually rich content, analogous to a traditional television programming guide. Topics tied to current events, trending cultural moments, breaking news, local happenings, and seasonal themes carry heightened importance on CTV. This contrasts sharply with mobile surfaces, where algorithmic personalization based primarily on historical engagement signals dominates. Meeting these expectations requires recommendation systems capable of incorporating real-world context to surface content that feels editorially curated yet remains personalized.

Existing recommendation systems, however, are not well suited to this challenge. Traditional multi-stage pipelines decompose the task into retrieval, ranking, and post-processing, where each stage consumes user  signals with a fixed schema. Incorporating novel contextual signals, e.g., cross-surface activity, requires engineering plumbing to propagate changes across multiple funnels. This approach is slow to adapt to new signal types and the system is expensive to maintain. 
Recent unified generative architectures seek to collapse this multi-stage pipeline into a single model. HSTU~\cite{zhaietal2024} replaces retrieval and ranking with a trillion-parameter sequential transducer; P5~\cite{geng2022recommendation} frames multiple recommendation tasks as text-to-text generation; and OneRec~\cite{deng2025onerecunifyingretrieverank,onerecthink} unifies retrieval and ranking through a single generative model.
However, these approaches still require structured behavioral sequences, while the heterogeneous contextual signals (such as trending topics, live events, breaking news, etc.) might come in varying formats, schema and frequencies.
Incorporating such signals still requires carefully designed retraining or bespoke feature engineering rather than relying on the built-in world knowledge of the large LLM to reason about the user journey cross-surface.

In this paper, we present an LLM-powered agentic recommendation system for content discovery on CTV that addresses these limitations. Our system employs LLM-based agents that autonomously reason over heterogeneous contextual signals expressed in natural language, including user histories, trending events, and cross-surface activity. Based on this reasoning, the agents produce structured decisions such as topic selection and content ranking. An orchestrator coordinates these agents alongside traditional ML components, allocating each task to the most suitable approach: LLM agents handle tasks that require contextual reasoning and open-ended generation, while traditional ML models handle latency-sensitive personalization tasks such as late-stage ranking, where they remain superior.

This agentic design enables the system to incorporate new signal types through prompt changes alone, without retraining models or building new feature pipelines.
This agentic design changes how contextual signals enter the recommendation process. Rather than requiring months of feature engineering and model retraining per signal type, the system can reason over any new information source as soon as it is expressed in natural language, bridging the gap between real-world context and personalized recommendation.
Our main contributions are as follows:
\begin{itemize}
    \item We describe the design of an agentic recommendation architecture for CTV that orchestrates LLM and traditional ML components, enabling flexible integration of diverse contextual signals.
    \item We present engineering solutions to productionization challenges, particularly inference latency.
    \item We share results from experiments and practical lessons learned from LLM-powered recommendation system.
\end{itemize}

\begin{figure*}[t] 
    \centering
    \includegraphics[width=0.95\textwidth]{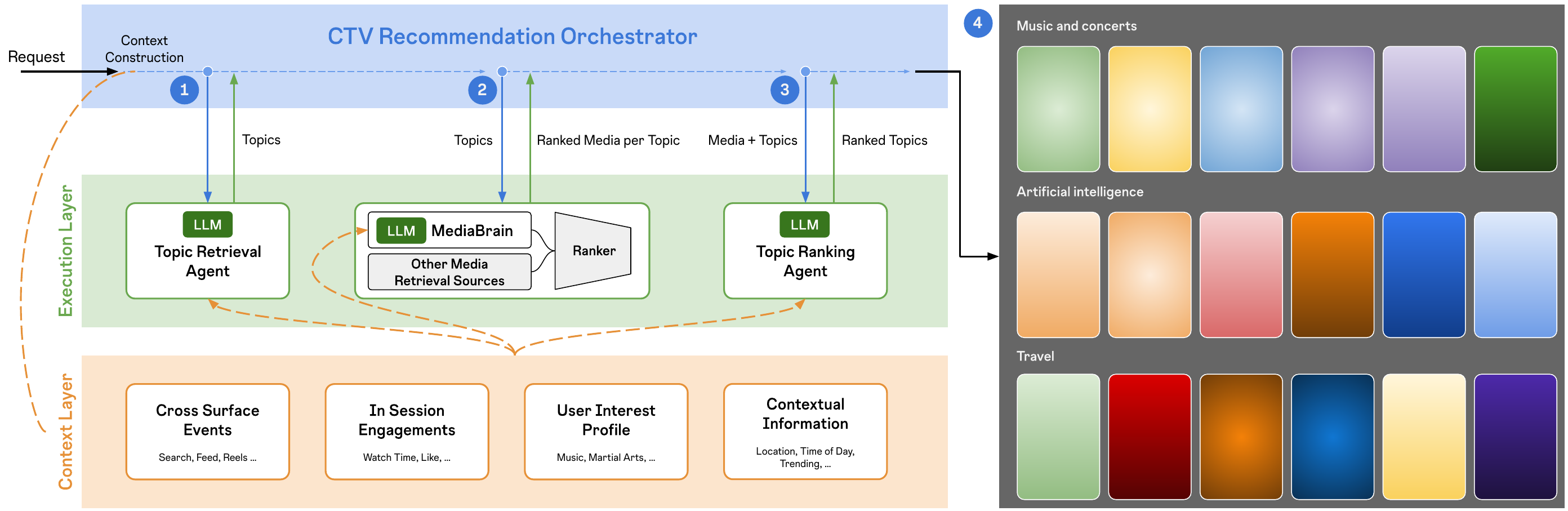}
    \caption{End-to-end workflow of the LLM-powered agentic recommendation system for connected TV (CTV) content discovery. An orchestrator coordinates three stages: (1) an LLM topic retrieval agent generates candidate topics; (2) a media retrieval and ranking agent, including MediaBrain, an LLM-native generative retrieval model, along with other retrieval sources, retrieves and then ranks media for each topic; and (3) an LLM topic ranking agent produces the final topic ordering. All components share access to a unified context layer. The output (4) is a personalized channel and media layout on the CTV discovery homepage. Media are organized into topic-based channels. Selecting a video initiates a chaining page experience.}
    \label{fig:agentic_ctv_flow}
\end{figure*}

\section{Background}

\paragraph{Limitations of existing recommendation architectures} Modern industrial recommendation systems, whether following the traditional multi-stage stack of candidate retrieval, ranking, and post-processing (e.g. diversity control), or adopting recent unified generative architectures such as OneRec~\cite{deng2025onerecunifyingretrieverank} that collapse these stages into a single model, share a common limitation: they are designed to consume structured behavioral signals with consistent schemas. While these architectures have proven highly effective for single-feed surfaces, they present challenges for incorporating heterogeneous, often unstructured contextual signals, such as trending topics, live events, or cross-surface user activity, which carry heightened importance in the CTV use case. Neither traditional multi-stage systems nor unified generative models possess the reasoning capability to naturally process information in varying formats and schemas. This rigidity makes it costly to adapt to new signal types and ingest contextual information, as it demands dedicated feature engineering, bespoke data ingestion pipelines, and carefully tuned heuristics. This has led to limitations in delivering a contextually rich CTV recommendation experience.

\paragraph{Leveraging LLMs for recommendation} Recent work has explored leveraging large language models to address these limitations \cite{ramanujam2026large,liang2026generative, Rajputetal2023,plum}. Comprehensive surveys~\cite{zhang2025surveylargelanguagemodel,peng2025surveyllmpoweredagentsrecommender} provide taxonomies of LLM-powered recommendation agents, categorizing approaches by how LLMs participate in the recommendation pipeline, as feature encoders, scoring functions, conversational interfaces, or autonomous agents. A recurring finding across these works is that while LLMs excel at understanding semantic content, interpreting user intent, and reasoning over heterogeneous information, they consistently underperform specialized collaborative filtering and deep learning models on ranking tasks that rely on behavioral signals at scale.

\paragraph{Agentic and multi-agent architectures} To bridge this gap, a growing body of work explores agentic architectures that decompose the recommendation process into specialized sub-tasks. \citet{huang2024recommenderaiagentintegrating} propose an early framework for integrating LLMs as interactive recommendation agents with tool-use capabilities. \citet{ranganathan2026multiagentvideorecommendersevolution} survey multi-agent patterns specifically for video recommendation, identifying common agent roles and orchestration strategies. \citet{zhang2026llmsorchestratorsconstraintcompliantmultiagent} frame the LLM as an orchestrator that coordinates multiple specialized agents under explicit constraints, while \citet{wu2026internalizingmultiagentreasoningaccurate} propose internalizing multi-agent reasoning into a single model for efficiency. \citet{chen2026memreccollaborativememoryaugmentedagentic} introduce memory-augmented collaborative agents that maintain persistent user representations across sessions. \citet{wang2026selfevolvingrecommendationsystemendtoend} demonstrate self-evolving recommendation systems at YouTube where LLM agents autonomously optimize model configurations end-to-end.

\paragraph{Evaluation and personalization} Robust evaluation of agentic recommender systems remains an open challenge. \citet{liu2026recoworldbuildingsimulatedenvironments} propose simulated environments for benchmarking agentic recommendation, while \citet{chauhan2026morebenchmarkingllmbased} find that simpler agent configurations can outperform complex ones, cautioning against unnecessary architectural complexity. \citet{xu2026personalizedllmpoweredagentsfoundations} survey foundations for personalized LLM agents, highlighting the tension between generalization and user-specific adaptation.

\paragraph{Gaps addressed by this work} Despite progress, several gaps remain. First, the majority of existing work operates in offline or simulated settings; reports of industry-grade agentic recommendation systems remain scarce. Second, most architectures target single-feed ranking tasks rather than the multi-carousel, topic-organized layout characteristic of CTV surfaces. Third, practical engineering challenges, particularly inference latency, for a smooth interactive TV experiences, receive limited attention in the literature. Our work addresses these gaps by presenting an industry-grade agentic system specifically designed for CTV content discovery, with detailed treatment of the engineering solutions that made real-time serving feasible.

\section{Approaches}

\subsection{Architecture Overview}

We propose an agentic architecture comprising an orchestration layer and three specialized agents, operating entirely off the user-facing critical path. 
Each agent is a self-contained component with an explicit input and output contract, orchestrated by a single layer that also owns context construction, producing a shared context object consumed by every downstream agent.
The orchestration layer aggregates information from a set of heterogeneous data sources containing structured, semi-structured, and unstructured data, such as cross-surface events, user in-session engagements, their long-term interest profile, and contextual information such as location and time. The insights and aggregated data are then assembled into a single context object in the form of natural-language statements for inclusion in downstream LLM prompts.

As shown in Figure~\ref{fig:agentic_ctv_flow}, a fresh user request reads the result from a precomputed cache. The agentic pipeline runs asynchronously after the response is delivered and writes back to the cache inventory for subsequent requests to consume. The orchestration layer performs \emph{context construction}, assembling a snapshot of the user's latest context and this shared context is then passed to three downstream agents that the orchestrator triggers in sequence: (1) the \emph{topic retrieval agent}, an LLM that consumes the context and emits a slate of new topic channel names for the page; (2) the \emph{media retrieval and ranking agent}, which dispatches each generated topic into the media retrieval infrastructure to produce per-topic candidate sets, and rank them; and (3) the \emph{topic ranking agent}, which ranks the resulting topic-with-media tuples for final delivery. Each agent is gated by configuration so that it can be enabled, disabled, or experimented on independently.

\begin{promptbox}{Example Context Object as Natural-Language}
Today is Wednesday, May 14, 2026.
Time of day: evening
User interest profile: Street Art, Urban Photography
Search: Tokyo Travel, Ramen Restaurants
Interaction: Sunset Timelapse, Drone Footage
Topics already watched this session: Coastal Sunsets, Street Food Around the World
Topics dismissed by user: Celebrity News
Topics with high engagement: Coastal Sunsets
Topics frequently skipped: Celebrity News, Daily Horoscope
Currently trending or seasonal events: NBA Playoffs 2026, Met Gala Highlights
User location context: New York
\end{promptbox}

\subsection{Topic Retrieval Agent}
The topic retrieval agent is implemented as a single LLM inference call to an instruction-tuned Llama-3 70B model~\cite{Dubeyetal2024} instance. The system prompt frames the model as a recommendation agent for a CTV surface and instructs it to return an array of short, specific, human-readable topics. The user prompt enumerates the seed topic names drawn from the user interest profile and embeds the context summary, ensuring the capture of both evergreen and timely user interest. The raw model output is then passed to two safety filters before being accepted by the downstream media retrieval and ranking agent: a curated sensitive topic blocklist and an asynchronous high-risk-content classifier.

\begin{promptboxhl}{Example Topic Retrieval Agent Prompt}
(*\textbf{System Prompt}*)
You are a content recommendation agent for Connected TV. Your job is to generate topic channel names that a user would enjoy watching based on their interests, session context, and what is happening in the world right now. Use your knowledge of current events, seasonal moments, trending cultural topics, and the current date and time to make your suggestions timely and relevant. Return an array of topic name strings. Each topic should be a short, specific, human-readable label. Generate diverse topics across different interest areas. Do NOT repeat topics the user has already watched this session.

(*\textbf{User Prompt}*)
The user has shown interest in these topics: Street Art, Urban Photography, Tokyo Travel, Ramen Restaurants, Sunset Timelapse, Drone Footage.

Context:
{context object from the orchestration layer}

Generate 10 new topic channel names for this user. Consider current events, seasonal relevance, and time of day when generating topics. Be specific and creative. Avoid generic labels.

\end{promptboxhl}

\subsection{Media Retrieval and Ranking agent}
The media retrieval and ranking agent dispatches each LLM-generated topic in parallel into the multi-stage recommendation pipeline, including MediaBrain -- a generative retrieval model with LLM backbone trained for CTV, followed by a lightweight early-stage ranking model that narrows the retrieval output to a few hundred candidates and a heavier late-stage ranking model that produces the final ordered slate. The ranking model follows the traditional multi-task multi-label (MTML) arch that scores and ranks candidates on engagement objectives such as watch time, likes, and shares to produce a personalized per-topic ordering. In this way, the system lets the LLM and traditional solutions each play to their strengths and complement each other. LLM-derived topics inherit scoring, integrity treatment, freshness gates, and inventory constraints without requiring the LLM to reason about media-level signals.

\subsection{Topic Ranking Agent}
The topic ranking agent determines the final ordering of topics rendered to the user. It consumes three inputs: the shared context object produced by the orchestration layer, topic-level engagement history summarizing how the user has historically interacted with each topic category, and the ranked media slates returned by the media retrieval and ranking agent. By combining contextual signals with topic engagement history and media quality indicators, the agent produces a ranking that balances user interest, topical diversity, and content freshness. Per-stage outputs at every agent boundary are persisted to a structured-log table, providing the provenance trail that grounds both online debugging and offline evaluation.

\begin{promptboxhl}{Example Topic Ranking Agent Prompt}
(*\textbf{System Prompt}*)
You are a topic ranking agent for Connected TV. Given a list of candidate topics with their engagement signals and media quality indicators, reorder them to maximize user satisfaction. Return an array of topic name strings in the recommended order. Balance user interest, topical diversity, content freshness, and timeliness.

(*\textbf{User Prompt}*)
Context:
{context object from the orchestration layer}

Candidate topics with signals:
1. "NBA Playoffs Highlights" - media_count: 24, avg_predicted_watch_time: 45s, user_topic_affinity: 0.82, freshness: 2h
2. "Hidden Tokyo Alleyways" - media_count: 18, avg_predicted_watch_time: 38s, user_topic_affinity: 0.91, freshness: 6h
3. "Met Gala 2026 Fashion" - media_count: 31, avg_predicted_watch_time: 28s, user_topic_affinity: 0.45, freshness: 1h
4. "Drone Cinematography Tips" - media_count: 12, avg_predicted_watch_time: 52s, user_topic_affinity: 0.88, freshness: 12h
5. "Street Ramen Reviews" - media_count: 15, avg_predicted_watch_time: 41s, user_topic_affinity: 0.79, freshness: 4h

Rank these topics for the user. Return an array of topic name strings in the recommended order.
\end{promptboxhl}

\subsection{Contextual Signals Consumption}
An advantage of this agentic architecture over traditional ML is the natural ingestion of heterogeneous contextual signals, whether structured, semi-structured, or unstructured. The orchestration layer ingests these signals, reasons through them to extract insights, and constructs the final context as natural-language statements in the LLM prompt for downstream agents.

For trending and seasonal events, e.g., music festivals, sports finals, cultural moments, we maintain a curated event feed populated by an offline editorial process, with each entry containing a name, brief description, geographic applicability, and an active date window. Rather than building dedicated feature transformations to embed these entries into the model, we serialize the active-event subset directly into the topic retrieval agent prompt. The agent natively incorporates relevant active events into its topic slate and stops surfacing them once their date window passes, providing built-in event recency and timelines. Onboarding a new event source requires only a feed extension, no model retraining is needed. For cross-surface and in-session activity, the context snapshot captures previously watched media, watched and dismissed topic names, dwell-time-by-topic, and skip counts. The topic retrieval agent is explicitly prompted to avoid repeating already-watched topics and is implicitly biased toward neighborhoods of high-engagement clusters. This loop produces feedback awareness across pages of the same session.

\subsection{Generative Retrieval}

While traditional ML retrieval models have served as the backbone of recommendation systems, these models often struggle with natural language understanding and are generally constrained by a fixed vocabulary.
In CTV, the topics generated by the topic retrieval agent are inherently open-ended. Inspired by recent advancements in generative retrieval~\cite{Tayetal2022,Rajputetal2023,plum,onerecthink}, to effectively process these topics and integrate diverse context objects, we developed MediaBrain: a generative retrieval model utilizing an LLM backbone to map natural language prompts directly to media indexed via semantic IDs (SIDs).

We leverage the media embedding, which is then quantized into a sequence of four-layer SID tokens via Residual-Quantization k-Means (RQ-kMeans)~\cite{Rajputetal2023}, with a codebook size of 1024 at each layer. The resulting tokens capture content distinctions in more granularity, from broad interest category (e.g., martial arts, cosmetics, sport) to narrower category representing niche interest (e.g., TV show, brand, celebrity). 

We initialize MediaBrain from a Llama 3.2 1B model~\cite{Dubeyetal2024}, expand its vocabulary with the SID codebook, and align SID tokens with natural language through supervised fine-tuning (SFT) on topic--media pairs. Inference proceeds in two stages: (1)~a \emph{token generation} step, in which the model autoregressively decodes a natural-language topic prompt into multiple SID token sequences via beam search; and (2)~a \emph{prefix lookup} step, which matches the generated SID prefixes against a pre-indexed SID-to-media mapping to retrieve a ranked set of candidate media for each topic.

\subsection{Productionization}

Preparing an LLM-powered agentic system to be industry-grade introduces engineering challenges that the academic literature seldom addresses. To start with, the dominant constraint is \emph{end-to-end latency}. The home screen must populate within hundreds of milliseconds for acceptable experience when user turns on the TV, which is equivalent to the forward inference pass of a small LLM alone. As a result, any naive placement of LLM inference on the hot request path immediately breaks the experience. 
Another constraint is \emph{throughput}. CTV surface is designed to serve large, geographically distributed traffic, the requests must sustain a reasonable QPS without saturating the underlying GPU tenant and leading to production scalability issues.

Our design removes real-time LLM inference from the critical path and enables it with a cache-based approach. An asynchronous post-processing stage fires after the user-facing response has been sent, running LLM inference in a non-blocking manner and writing the results to the cache. Subsequent requests then read from this cache, where served topics and candidate sets are looked up rather than computed online.

To better balance freshness and latency, we adopt a multi-layer, per-component caching strategy that allows each stage to cache at its own pace: more static sources, such as the long-term interest profile, can be cached aggressively, while faster-evolving signals, such as in-session activity, are fetched fresh most of the time. By decoupling cache lifetimes per component, only the stages whose inputs have materially changed need to be re-triggered on each refill cycle, maximizing latency savings while preserving result quality.

While this strategy applies broadly across the agentic pipeline, optimizations were applied at every layer of the stack. Given space constraints, we defer the detailed treatment of MediaBrain, the generative-retrieval component used by the media retrieval and ranking agent, to the Evaluation section, where we present both the engineering optimizations and the relevance improvements as a unified case study.
          
\section{Evaluation and Results}

As noted in the preceding section, optimizations were applied across every layer of the agentic pipeline. In this section, we focus on MediaBrain, the generative-retrieval component within the media retrieval and ranking agent, as a representative case study. We first present the serving performance optimizations that made it industry-grade, then evaluate topic and media retrieval quality through evaluations and experiments.

\subsection{Serving Performance}
\subsubsection{Per-component caching} 
MediaBrain online inference ordinarily takes approximately 500ms for token generation and an additional 150ms for the prefix lookup. Because beam search produces near-deterministic, unique outputs when the prompt is held fixed, with strictly higher cumulative sequence log-probability than temperature-controlled sampling, the generated tokens can be cached and reused directly, eliminating the 500ms generation latency on subsequent requests and making cached tokens safe to reuse across chaining pages and discovery homepage within the same session without quality drift. The prefix lookup result is cached separately, keyed by the generated token set and the query-time scoring parameters, so that the same prefix can be looked up with different scoring configurations and return different candidate sets without re-invoking the LLM. Together, these per-component caches allow MediaBrain to achieve neutral user-facing latency despite running a billion-parameter LLM in the system.

\subsubsection{Chained retrieval via cursor-anchored persistence} 
The per-component cache described above eliminates the 500ms token generation latency when the asynchronous refill path has already warmed the cache. However, the very first chaining request in a session, and any subsequent page that arrives before the refill completes, would still incur the full generation latency under a naive implementation that re-invokes the LLM for every page. To avoid this, we serialize the generated token set directly into every chaining response payload such that the client can include it in the next chaining request. In this way, the server can simply extract tokens from subsequent chaining requests and proceed straight to the prefix lookup, without needing to regenerate it every time, saving the 500ms generation latency on every page after the first. This approach is logically equivalent to memoizing LLM output by topic query. However, using the cursor as the carrier removes the need for a server-side per-session cache and naturally bounds the lifetime of cached tokens to the session in which they were generated.

\subsubsection{Throughput optimization} 
While the beam search can help generate more diverse and relevant candidate SIDs than temperature sampling, it is expensive, where the time complexity is proportional to beam width and output token count. For our configuration of a small beam over 4-layer SIDs, this expands to roughly 150 vLLM decoding calls per query. As a result, a naive configuration on a generic vLLM~\cite{vllm} serving stack achieved only 2 QPS per A100 card, far below the expected level. 

Through our optimization, we firstly enabled concurrent request handling at the serving-framework level to resolve Python-side serialization bottlenecks, resulting in a 20 times throughput increase. Secondly, we identify and fix the CPU bottleneck in the beam search implementation of vLLM engine, including re-tokenizing the system prompt for each decoding call and deep-copying sampling parameters for each decoding call, raising performance to 80 QPS. Finally, we identified the GIL saturation by distributing beam search orchestration across multiple vLLM model instances, even on the same card. This increased GPU utilization by 4 times and pushed throughput to 200 QPS. Together, these optimizations provided a cumulative 100 times speedup, making the MediaBrain model viable for industry-grade serving.

\subsection{Retrieval Quality}
\subsubsection{Media Retrieval: MediaBrain}
Media retrieval relevance determines whether the topics generated by the upstream agent translate into a coherent user experience. To measure this, we conduct offline relevance evaluation using an LLM judge that scores each retrieved media as topically relevant or not. We construct three evaluation benchmarks at different levels of topic popularity:
\begin{enumerate}
    \item \textbf{CTV100 / CTV1000.} The top 100 and top 1000 topics by impression volume, representing the head distribution that covers the majority of user requests.
    \item \textbf{Niche Entities.} A hand-curated set of 150 entity-driven topics---cities, celebrities, and brands---that test the model's ability to retrieve relevant media for personalized, long-tail queries.
\end{enumerate}

\begin{figure}
  \centering
  \includegraphics[width=0.6\linewidth]{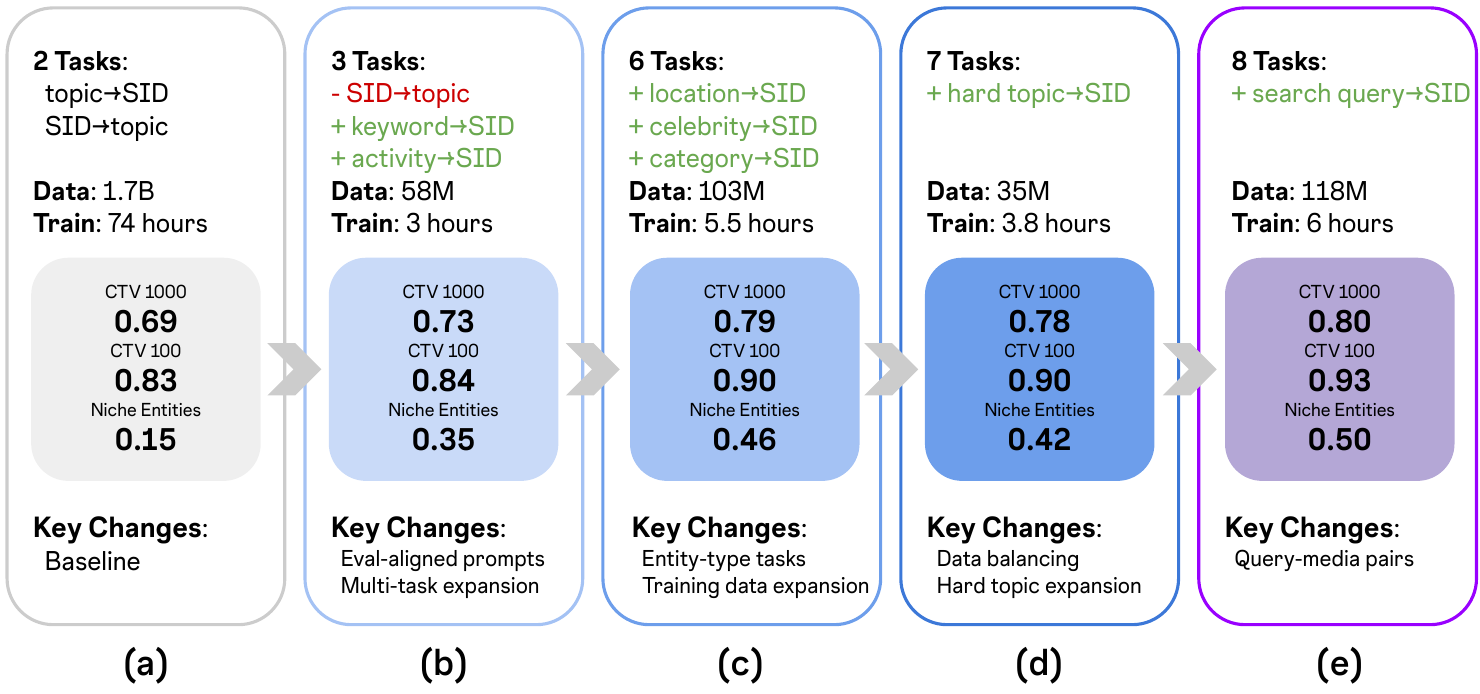}
  \caption{Media retrieval relevance results for MediaBrain across training data iterations for three benchmarks. Notably the difficult niche entity topics improve from $0.15$ to $0.50$.}
  \label{fig:instabrain-iterations}
\end{figure}

Through our modeling, we identify training data quality to be the most important.
We start with multiple annotation pipelines as data sources. We firstly extract CTV-relevant information to construct diverse training tasks. We then filter noisy samples to retain only media with salient visual patterns, improving the signal-to-noise ratio and training efficiency. Our training tasks are summarized in Table~\ref{tab:training_data}.
Figure~\ref{fig:instabrain-iterations} illustrates the topical relevance improvement over the iterations with the key changes at each iteration.

\begin{table}[htbp]
  \caption{Training tasks for MediaBrain alignment. Task diversity prevents model from overfitting to a specific pattern.~\cite{AllenZhu-icml2024-tutorial}}
  \label{tab:training_data}
  \centering
  \begin{tabular}{ll}
    \toprule
    \textbf{Task} & \textbf{Task Example} \\
    \midrule
    activity $\rightarrow$ SID     & watching reels, playing soccer \\
    location $\rightarrow$ SID     & Tokyo, Brooklyn, Bali \\
    celebrity $\rightarrow$ SID    & Taylor Swift, LeBron James, Zendaya \\
    Brand $\rightarrow$ SID   & Hello Kitty, Lululemon \\
    search query $\rightarrow$ SID & modern minimalist design ideas,  \\
    & cook ramen at home \\
    \bottomrule
  \end{tabular}
\end{table} 

Moreover, we found that the model can handle open-ended topics passed from upstream agent, and prompt optimization for different aesthetics, platform-surface, and diversity. This property ensures we are able to inject context to prompt without model retraining, to make the model adjust for different topics/styles from the upstream agent.
When prompted with ``Suggest a media about Interior Design with LED lighting'', the model retrieves content centered on LED signage fabrication and neon lighting installations. One root cause is from the training and inference discrepancy. The training data contains detailed multi-sentence descriptions (e.g. interior design featuring modern minimalist aesthetics with natural lighting and sustainable materials), while CTV inference prompts use only short topics (e.g. Interior Design). The model trained on verbose descriptions generates SIDs biased toward whichever narrow subcategory dominates the training distribution. In inference, this bias is exposed when prompting with a short topic, leading to low topic relevance and bad user experience.
Without re-training the model, we are able to refine the result by injecting additional context into the prompt as ``Suggest a media about interior design for creating a pleasing living environment with modern minimalist wooden aesthetics''. The retrieved results shift accordingly, surfacing content featuring warm-toned, wood-accented living spaces with minimalist furniture and ambient lighting. Finally, when the prompt introduces a platform-specific context as ``Suggest a media about interior design that is suitable for the social media platform'', the model returns visually curated content that blends interior design inspiration with product-oriented framing. 
These examples demonstrate that the model, after training with SID alignment tasks, can follow open-ended instructions from an upstream agent and adjust its retrieval behavior across different aesthetic styles and platform requirements without model retraining. This flexibility ensures that contextual signals can be injected directly into the prompt at serving time.

\section{Discussion and Future Work} 

\begin{figure}
  \centering
  \begin{subfigure}[b]{1\linewidth}
    \centering
    \includegraphics[width=0.6\linewidth]{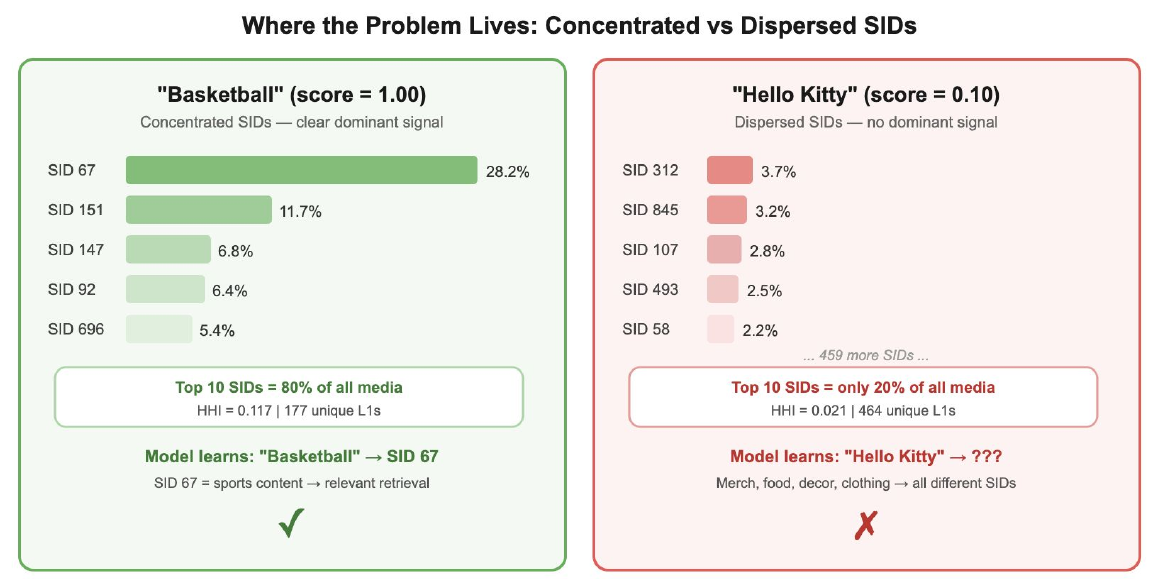}
    \caption{L1 SID distribution for a concentrated topic (\emph{Basketball}, left) versus a dispersed topic (\emph{Hello Kitty}, right).}
    \label{fig:sid-qualitative}
  \end{subfigure}

  \begin{subfigure}[b]{1\linewidth}
    \centering
    \includegraphics[width=0.6\linewidth]{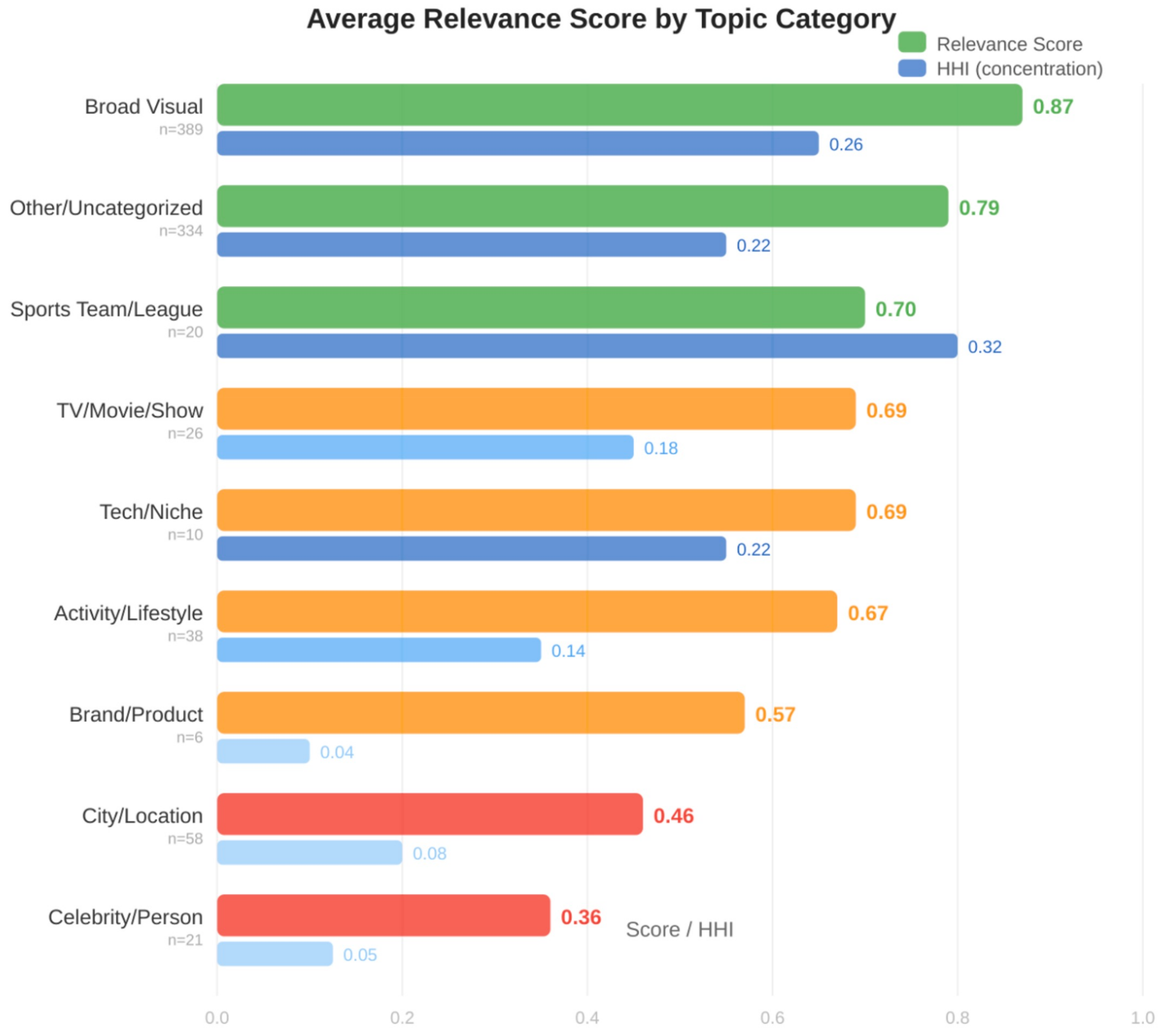}
    \caption{Average relevance score and HHI concentration by topic category. Entity-driven categories (Celebrity/Person, City/Location, Brand/Product) exhibit the lowest concentration and relevance.}
    \label{fig:sid-hhi}
  \end{subfigure}

  \caption{SID topic dispersion vs. performance of each topic.}
  \label{fig:sid_bottleneck}
\end{figure}

\subsection{Closed-loop Feedback from Agents}

The current architecture is a unidirectional pipeline: the orchestration layer constructs context, and agents consume it in sequence without propagating quality signals back upstream. A key limitation of this design is that failures at one stage are invisible to earlier stages. For example, if the topic retrieval agent consistently generates topics for which the media retrieval and ranking agent finds poor-quality or insufficient inventory, the topic agent has no mechanism to learn from this outcome and adjust its generation strategy. Similarly, if the topic ranking agent observes that certain topic categories are repeatedly skipped by users, this signal does not currently flow back to inform topic retrieval in future sessions. We envision extending the architecture with a closed-loop feedback mechanism, where downstream agents emit structured quality signals - such as per-topic media coverage, media relevance scores, and user engagement outcomes - that are persisted to a cross-session memory layer. This memory would be distinct from the in-session context: rather than capturing what the user is doing now, it would accumulate what has worked and what has not across sessions, specifically to refine agent behavior. The orchestration layer would inject relevant memory entries into agent prompts, enabling the topic retrieval agent to avoid generating topics with historically poor media coverage, and enabling the topic ranking agent to down-weight topic categories that a user has consistently ignored. This design transforms the pipeline from a stateless, session-scoped system into one that continuously improves its recommendations through agent-level learning, without requiring model retraining.

\subsection{SID Bottleneck for Niche Entities}

We observe that for MediaBrain, relevance gains saturate for niche entities even after we augment training data. E.g., niche topic (e.g., hello kitty) is harder to be improved than general topic (e.g., basketball).
We hypothesized that this might be caused by the SID bottleneck not being able to capture very fine-grained concepts.
To quantify it, we introduce \emph{SID-topic dispersion}.
Given a topic, the corresponding media with this topic can have $N$ different layer 1 (L1) SID tokens. Let $s_i$ denote the percentage of media within cluster $i$ and $\sum_{i} s_i = 1$.
We use \emph{Herfindahl-Hirschman Index} (HHI) $\sum_{i=1}^{N} s_i^2$ to measure the dispersion of these clusters. Its values are between $1/N$ and $1$, where $1/N$ means uniformly distributed (maximally dispersed), while $1$ means all media falls in a single cluster (maximally concentrated). 

We compare across 902 topics with different relevance performance (Figure~\ref{fig:sid_bottleneck}). We find that there is 0.25 Spearman-$\rho$ correlation between SID-topic dispersion and topic relevance, indicating that some of the hard topics are caused by the SID bottleneck. 
We hypothesize this bottleneck could come from the media embedding as well as the quantization process.
In the future work, we are leveraging HHI as a fast check on SID quality, given they are able to predict per-topic learnability before expensive training. 

\section*{Acknowledgement}
We gratefully acknowledge discussion and support from Xiyuan Wang, Chen Yuan, Jun Xiao, Qifei Wang, Xiangjun Fan, Benyu Zhang, Rui Li, Lihan Bin, Kyle Yan, Hang Cui, Geoffrey Wang, Angela Wise.

\clearpage
\newpage
\bibliographystyle{plainnat}
\bibliography{paper}

\end{document}